\documentclass[twocolumn,superscriptaddress,showpacs,pra]{revtex4}
\usepackage{epsfig}
\usepackage{epstopdf}
\usepackage{delarray}
\usepackage{amsmath, amssymb}
\usepackage{bm}
\usepackage{graphicx}
\usepackage{placeins}
\usepackage{color}

\begin{document}
\title{Rogue Heat and Diffusion Waves}

\author{Cihan Bay\i nd\i r}
\email{cbayindir@itu.edu.tr}
\affiliation{Associate Professor, Engineering Faculty,  \.{I}stanbul Technical University, 34469 Maslak, \.{I}stanbul, Turkey \\
						 Adjunct Professor, Engineering Faculty, Bo\u{g}azi\c{c}i University, 34342 Bebek, \.{I}stanbul, Turkey \\
						 International Collaboration Board Member, CERN, CH-1211 Geneva 23, Switzerland}

\begin{abstract}
In this paper, we numerically show and discuss the existence and characteristics of rogue heat and diffusion waves. More specifically, we use two different nonlinear heat (diffusion) models and show that modulation instability leads to generation of unexpected and large fluctuations in the frame of these models. These fluctuations can be named as rogue heat (diffusion) waves. We discuss the properties and statistics of such rogue waves. Our results can find many important applications in many branches such as the nonlinear heat transfer, turbulence, financial mathematics, chemical or biological diffusion, nuclear reactions, subsurface water infiltration and pore water pressure diffusion modeled in the frame of nonlinear Terzaghi consolidation models, just to name a few.

\end{abstract}
\maketitle

\section{\label{sec:level1}Introduction} 
One of the most widely and comprehensively studied models in physics, mathematics and engineering is the heat equation. Heat equation, also known as the diffusion equation, is commonly used in modeling diverse phenomena in these branches of science. Heat (diffusion) equation, first developed and solved by Joseph Fourier in 1822 \cite{Fourier1822, Fourier1878}, was introduced to the scientific literature as a model to analyze the distribution of a quantity, such as heat, over time in a medium. Today, the existing literature on this subject is vast \cite{Nkashama1984, Cazenave2001, Varlamov2000, Cazenave2011, Vazquez2017, Aronson, Vazquez1987, Vazquez2003, Vazquez2004, Vazquez2007book, Bernis_Hulshof_Vazquez,  Aronson_Vazquez, Kamin_Vazquez, Gurtin77, Aronson_Weinberger, Vazquez2014, Vazquez2015}.

The majority of equations of heat (diffusion) type proposed in the literature are used to model the diffusion of heat and chemicals, however, they have a wider range of possible applications. Diffusion type equations are commonly used in modeling Brownian motion and Kolmogorov diffusion \cite{Bramson78, Bramson83} of the statistical phenomena in financial mathematics and topology. One of their possible usage areas is the modeling of the propagation of turbulent burst and shock waves \cite{Kamin_Vazquez, LandauLifshitz, RaizerZeldovich} and nuclear reactor physics \cite{Masterson}. Diffusion of biological populations and nerves are also analyzed within the frame of nonlinear diffusion equations \cite{Gurtin77,  Aronson_Weinberger, Aronson_Weinberger1978}. Richard's equation, which is used for modeling the diffusion of the subsurface water, is a nonlinear diffusion type equation \cite{Chow}. Additionally, it is known that the Terzaghi's theory of consolidation can be formulated in terms of the linear and nonlinear heat equations, which are used as models to analyze the diffusion of excessive pore water pressure and settlement in the foundations of structures \cite{HoltzKovacs}. This list is brief and one can come up with many different possible application areas of the heat (diffusion) equation after a more comprehensive literature survey.

Throughout the historical development of heat equation, the earliest forms of the equations studied was linear and the vast majority of the studies and engineering designs still rely on the linear theory, as discussed in various textbooks such as \cite{Greenberg}. However, many studies on the different forms of nonlinear heat (diffusion) equation types do also exist in the literature including but are limited to those discussed in \cite{Vazquez1987, Nkashama1984, Cazenave2001, Varlamov2000, Cazenave2011, Vazquez2017, Aronson, Vazquez2003, Vazquez2004, Vazquez2007book, Vazquez2014, Vazquez2015, Bernis_Hulshof_Vazquez, Aronson_Vazquez}. Researchers have studied many different forms of equations having different nonlinearities, such as the porous medium equation and equations having fractional order derivatives in these papers.

On the other hand, it is known that nonlinear wave equations exhibit an important and fascinating phenomena called rogue waves \cite{Kharif, Peregrine, Akhmediev2009a, Akhmediev2009b, Akhmediev2011, Wang, Zhao2013, dqiu}. In the ocean engineering and fiber optics literature, the rogue waves are defined as the waves having a height more than 2 times the significant wave height in a chaotic wavefield which has many spectral components and peaks \cite{Kharif, Peregrine, Akhmediev2009a, Akhmediev2009b, Akhmediev2011}. Additionally, rogue waves have a very dynamic behavior in the wavefield, they are unexpected, that is `they appear from nowhere and disappears without a trace' \cite{Akhmediev2009a, Akhmediev2009b, Akhmediev2011}. They are observed in hydrodynamics \cite{Kharif}, optics and quantum mechanics \cite{Akhmediev2009b, Akhmediev2011, Birkholz, Bay_Zeno}, Bose-Einstein condensation \cite{Akhmediev2009a, Akhmediev2009b, Akhmediev2011}, just to name a few. Such rogue waves are generally studied within the frame of the nonlinear Schrödinger equation (NLSE) or NLSE like equations  \cite{Akhmediev2009a, Akhmediev2009b, Akhmediev2011, Soto2014RwSSchaotic, BayPRE1}  or fully nonlinear wave equations such as the one discussed in \cite{Baysci}. 

Rogue waves can be very catastrophic for marine travel and engineering, thus, they generally needed to be avoided in the marine environment for the safety and cost effectiveness of the marine operations. Two examples of their catastrophic results in the marine environment are the damaging of the Draupner platform and the sinking of the cargo vessel El Faro.  Although they possess great danger in the marine environment, in fiber optics they are generally desired to satisfy certain energy thresholds and for locating the information when techniques like matched filtering or compressive sensing are utilized. With these motivations, their early detection became an active area of research and researchers have proposed different spectral and nonlinear time series analysis techniques for their early detection\cite{Akhmediev2011, BayPLA, BayPRE2, Birkholz}.

In this paper, we numerically prove that rogue heat and diffusion waves can exist in the nonlinear media. With this motivation, we consider two different nonlinear heat (diffusion) equation types. We solve these model equations numerically using a Fourier spectral scheme in which FFTs are used to evaluate the spatial derivatives. The temporal derivatives are handled by a $4th$ order Runge-Kutta time stepping scheme. We show that noise imposed on the dependent parameter U, which represents temperature (concentration), triggers modulation instability (MI), which turns the initial conditions into a chaotic ones. In these chaotic fields, heat and diffusion waves having anormally large amplitudes and having a very dynamic nature can be observed. We name such waves as rogue heat (diffusion) waves, and redefine them within the frame of the heat (diffusion) equation. Additionally, the discuss the formation probabilities of such rogue heat waves within the frame of the two different models considered. We comment on our findings and discuss the possible usage and limitations of our results. We also comment on possible future research directions for our findings.

\section{\label{sec:level2}Methodology} 
First of the two nonlinear heat (diffusion) models considered in this paper can be given as
\begin{equation}
U_t = (U^{m-1}U_x)_x + \epsilon(x,t)
\label{eq01}
\end{equation}
where various $m$ values satisfying the $m>1$ condition is used to model the different degree of nonlinearities \cite{LandauLifshitz, RaizerZeldovich, Esteban1988}. In here, the last parameter represents the noisy input and it is taken as $\epsilon(x,t)=av(x,t)$ where $a=0.05$, throughout this paper. $v(x,t)$ is the set of uniformly distributed random numbers generated in the interval of $[-1,1]$ at each time step. Such a selection for the range of random numbers ensures that the mean of the input is zero, so that no net energy is pumped into the system during time evolution. For $\epsilon(x,t)=0$, the equation reduces to the regular nonlinear heat equation, which is used to study turbulent bursts and shock waves in some of the literature \cite{LandauLifshitz, RaizerZeldovich, Esteban1988}. We will refer to the model equation given by Eq.(\ref{eq01}) as the Model 1 throughout this paper. 
The second model considered in our paper is the porous medium equation which can be written as
\begin{equation}
U_t = \Delta U^m + \epsilon(x,t)
\label{eq02}
\end{equation}
where $m$ and $\epsilon(x,t)$ are as before. Compared to Model 1, the porous medium equation has been more comprehensively studied \cite{Aronson, Vazquez2003, Vazquez2004, Vazquez2007book}. Porous medium equation is commonly used as a model for gases in porous media, high energy physics, population dynamics and underground infiltration, just to name a few of its uses \cite{Aronson, Vazquez2003, Vazquez2004, Vazquez2007book}. We will refer to the porous medium equation given by Eq.(\ref{eq02}) as the Model 2 in this paper. In both of these models, $U(x,t)$ is the space and time dependent function. Depending on the phenomena modeled, $U$ can represent the temperature, concentration, population, etc.

Starting from the initial conditions, the time stepping of these two model equations can be performed using a $4^{th}$ order Runge-Kutta scheme. The four slopes of the $4^{th}$ order Runge-Kutta scheme can be calculated at each time step as
\begin{equation}
\begin{split}
& s_1=h(U^n, t^n, x) \\
& s_2=h(U^n+0.5 s_1dt, t^n+0.5dt, x) \\
& s_3=h(U^n+0.5 s_2dt, t^n+0.5dt, x) \\
& s_4=h(U^n+s_1dt, t^n+dt, x) \\
\end{split}
\label{eq03}
\end{equation}
where $n$ shows the iteration count and $dt$ is the time step which is selected as $dt=2 \times 10^{-4}$ throughout this study. The function $h$ denotes the right-hand-side of the model equation given either by Eq.(\ref{eq01}) or Eq.(\ref{eq02}).  Then at each time step, the values of the independent time parameter and dependent $U$ parameter can be computed using these slopes as 
\begin{equation}
\begin{split}
& U^{n+1}=U^{n}+(s_1+2s_2+2s_3+s_4)/6 \\
& t^{n+1}=t^n+dt\\
\end{split}
\label{eq04}
\end{equation}
Spatial derivatives in these models are handled spectrally using the FTT routines. The first term on the right-hand-side of Model 1 is numerically calculated as 
\begin{equation}
\frac{\partial U^{m-1}U_x}{\partial x} = F^{-1} \left[ ik F \left[ U^{m-1} F^{-1} \left[ikF[U] \right] \right] \right]
\label{eq05}
\end{equation}
where $F$ and $F^{-1}$ shows the Fourier and the inverse Fourier transforms, respectively, $i$ shows the imaginary unity and $k$ is the wavenumber vector which has exact $N$ multiples of the fundamental wavenumber, $k_o=2 \pi/(2L)$. In here, $L$ is the half of the spatial domain length and taken as $L=100$. Similarly, the first term on the right-hand-side of Model 2 can be numerically calculated using
\begin{equation}
\frac{\partial^2 U^m}{\partial x^2} = F^{-1} \left[ -k^2 F[U^m] \right]
\label{eq06}
\end{equation}
where the notations of $F, F^{-1}, i, k, L$ is as before. The number of spectral components is selected as $N=512$ for the efficient computations of FFTs. The nonlinear products are calculated by simple multiplication in the spatial domain. In our simulations, we solve both of the model equations using the scheme discussed above. The initial condition for Model 1 is taken as $U_0=5+e^{-0.01(x-L)^2}$. Similarly, the initial condition for Model 2 is taken as $U_0=10+e^{-0.01(x-L)^2}$.

\section{\label{sec:level3}Results and Discussion}
 
\begin{figure}[htb!]
\begin{center}
\hspace*{-0.7cm}
   \includegraphics[width=3.6in]{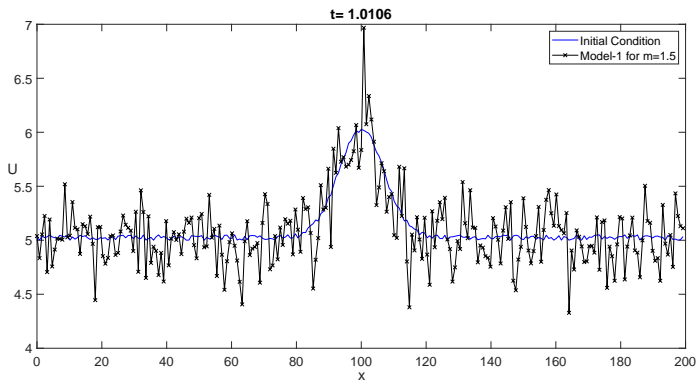}
  \end{center}
\caption{\small A snapshot of the chaotic field generated in the frame of Model 1 for m=1.5.}
  \label{fig1}
\end{figure}

A typical chaotic heat wave field is depicted in Fig.~\ref{fig1}. This chaotic wave field is generated in the frame for Model 1 for which the nonlinearity parameter is selected as $m=1.5$. The initial condition for this simulation is $U_0=5+e^{-0.01(x-L)^2}$ with a uniformly distributed random noise produced by the expression $\epsilon(x,t)=0.05 v(x,t)$ is superimposed on it, as discussed above. This initial condition is depicted in blue in Fig.~\ref{fig1}. Then, starting from this initial condition the numerical solution is obtained using the spectral scheme with $4^{th}$ order Runge-Kutta time integrator. Random noise with the same amplitude of $0.05$ is applied iteratively at each time step. However, as Fig.~\ref{fig1} suggests, such a small random noise triggers MI and large fluctuations are formed in the chaotic wavefield within few time steps.
Such fluctuations have the similar characteristics of rogue waves. Firstly, they are changing very rapidly and have a very dynamic wavy behavior. Secondly, they have significantly larger amplitude deviations compared to the noise injected into the system. Thus, we can possibly name such waves as the rogue heat (diffusion) waves. The precise mathematical criteria for classifying such a heat (diffusion) wave as a rogue heat (diffusion) wave remains as an open question and a widely accepted criteria would be needed. Rogue waves are classified as waves having a height at least two times the significant wave height in a chaotic field studied in the frame of the wave equations such as the NLSE \cite{Akhmediev2009a, Akhmediev2009b, Akhmediev2011}. However, the nature of the heat (diffusion) type of equations is different than wave equations. It is well known that unsteady parts of the solutions decay in time due to the effect of the diffusion. Therefore, the wavefields generated in the frame of the heat equations generally will not reach a steady state. Therefore, the criteria for heat waves being rogue should be a time dependent one. One possible solution to overcome this definition problem can be using some confidence levels. Linear heat equation can be used for this purpose. The solution of the nonlinear heat equation having such large fluctuation and solution of the linear equation, $U_{lin}$ can be done simultaneously. Since the MI is triggered due to nonlinear terms, in a typical run the linear solution $U_{lin}$ would start as the blue line depicted in Fig.~\ref{fig1} and would decay smoothly without exhibiting large fluctuations. Using the linear solution $U_{lin}$ as the threshold criteria, it is possible to classify large fluctuations with amplitudes of $U>1.25 U_{lin}$ or $U<0.75 U_{lin}$ as rogue heat (diffusion) peak and dip waves. This definition and confidence levels are vague, researchers may offer different criteria for the physical meaningful analysis of the phenomenon they investigate.

\begin{figure}[htb!]
\begin{center}
\hspace*{-0.7cm}
   \includegraphics[width=3.6in]{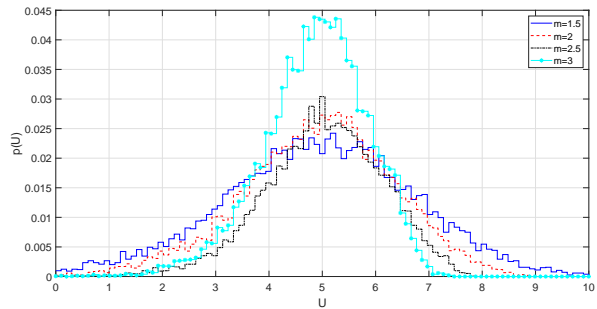}
  \end{center}
\caption{\small Probability of occurrence of heat (diffusion) fluctuation amplitudes for Model 1.}
  \label{fig2}
\end{figure}

In Fig.~\ref{fig2}, we depict the probability of occurrence of heat (diffusion) fluctuation amplitudes in the frame of Model 1 for different values of the nonlinearity parameter, $m$. Each of these probability distributions functions (PDFs) include more than 10000 waves recorded at different times of temporal evolution, where the same times of $t=10, 12, 14, 16, 18$ are used to capture different chaotic heat (diffusion) fields modeled using different $m$. Although generally the larger values of $m$ lead to stronger MI thus more rogue waves, the constant used in the initial condition suppresses this effect since it becomes significantly larger when its $(m-1)^{th}$ power is taken.

\begin{figure}[htb!]
\begin{center}
\hspace*{-0.7cm}
   \includegraphics[width=3.6in]{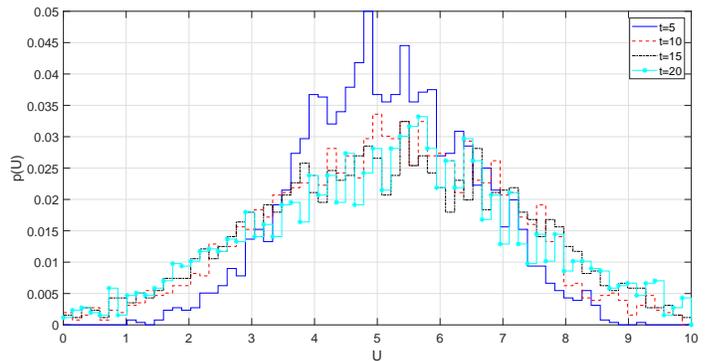}
  \end{center}
\caption{\small Probability of occurrence of heat (diffusion) fluctuation amplitudes for Model 1 with $m=1.5$ at different times.}
  \label{fig3}
\end{figure}
In Fig.~\ref{fig3}, we depict the probability of occurrence of heat (diffusion) fluctuation amplitudes in the frame of Model 1 using $m=1.5$ for different times. The behavior of the curves in Fig.~\ref{fig3} indicates that as time increases, the probability of rogue heat (diffusion) wave occurrence increases initially, after some adjustment takes place. However, after time gets larger, due to the effect of diffusion the PDFs for larger times first become almost indistinguishable, suggesting the balancing of nonlinear interactions and diffusion. For longer temporal evolutions, the PDFs are expected to exhibit less occurrence of large fluctuation amplitudes. Additionally, one can observe that the shapes of PDFs obtained at larger times have a more asymmetric structure and have right-skewness. This is clearly the effect of decay in amplitudes due to diffusion. The results in Fig.~\ref{fig2} and Fig.~\ref{fig3} supports that for the selected parameters, $U$ can attain values in the interval of approximately $[0,10]$ within the frame of Model 1.

\begin{figure}[htb!]
\begin{center}
\hspace*{-0.7cm}
   \includegraphics[width=3.6in]{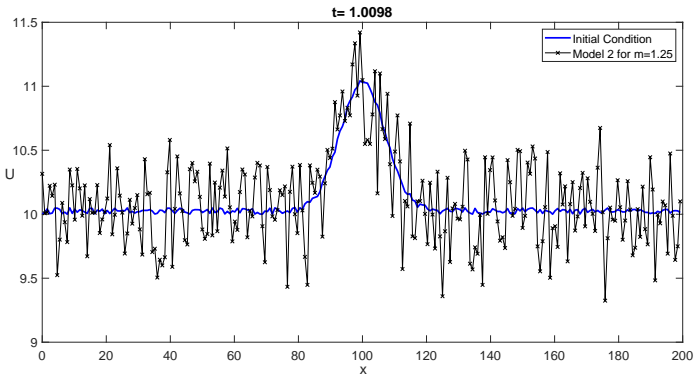}
  \end{center}
\caption{\small A snapshot of the chaotic field generated in the frame of Model 2 for m=1.25.}
  \label{fig4}
\end{figure}

Next, we turn our attention to Model 2 (porous medium equation) and repeat the similar numerical analysis. In Fig.~\ref{fig4} we depict a typical chaotic heat (diffusion) field generated in the frame of this second model. In our simulations we observe that for larger values of $m$, the numerical scheme can get unstable, therefore smaller $dt$ would be needed when larger values of $m$ are used. As one can see in Fig.~\ref{fig4}, the noise injected into the systems triggers MI again thus rogue heat (diffusion) waves are formed in the frame of the Model 2 (porous medium equation). Their dynamics are also similar to those of the Model 1, as their amplitude rapidly oscillates about the mean initial condition. 

\begin{figure}[htb!]
\begin{center}
\hspace*{-0.7cm}
   \includegraphics[width=3.6in]{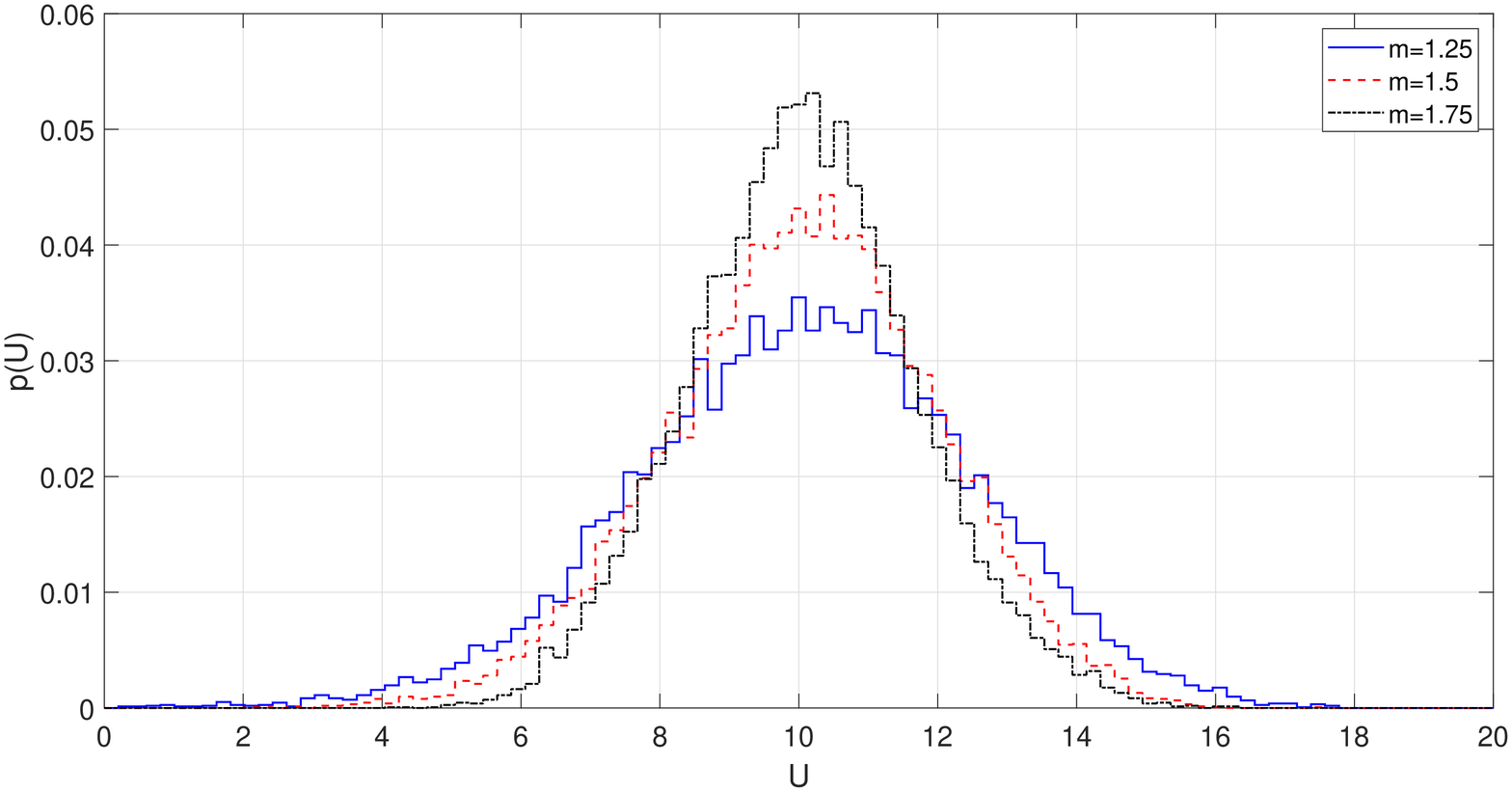}
  \end{center}
\caption{\small Probability of occurrence of heat (diffusion) fluctuation amplitudes for Model 2.}
  \label{fig5}
\end{figure}

\begin{figure}[htb!]
\begin{center}
\hspace*{-0.7cm}
   \includegraphics[width=3.6in]{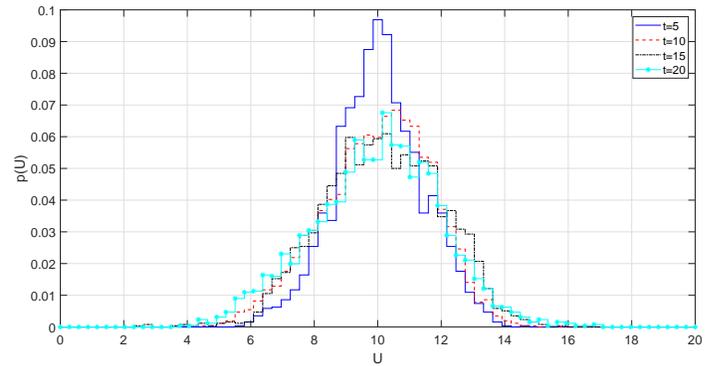}
  \end{center}
\caption{\small Probability of occurrence of heat (diffusion) fluctuation amplitudes for Model 1 with $m=1.25$ at different times.}
  \label{fig6}
\end{figure}

As the PDFs plotted in Fig.~\ref{fig5} and Fig.~\ref{fig6} indicates $U$ can attain values in the interval of approximately $[0,20]$ within the frame of Model 2, when the selected parameters are used. The characteristics of PDFs plotted in these figures are similar to the ones discussed above for Model 1. The difference is that, larger values of rogue heat (diffusion) wave amplitudes can be observed in the frame of the Model 2 (porous medium equation)  due to stronger nonlinearity.

Additionally, we have investigated the effects of noisy diffusion constant on rogue heat (diffusion) wave generation. For this purpose, we have used a more general version of Model 1, which can be given as $U_t = (D(x,t) U^{m-1}U_x)_x$. Using a diffusion parameter in the form $D(x,t)=c+av(x,t)$, where $c$ is a constant and other parameters are as before, we have observed that the superimposed noise remains small and MI does not trigger the occurrence of the rogue heat (diffusion) waves for such a scenario.

The analysis and the framework proposed in our paper can shed light upon to many different future directions. First of all, our results can be generalized to other types of nonlinear heat (diffusion) equations having different order nonlinearities and fractional order derivatives. The analysis of MI and rogue heat (diffusion) wave generation for those type of models can lead to important contributions to the literature. Additionally, our results are expected to have many important applications in science and engineering. The two models used in our study are nonlinear heat (diffusion) models with source terms. Therefore, the dynamics of such models under the effect of noise can be investigated by our approach and occurrence of rogue heat (diffusion) waves can be observed in similar models. One practical example is the light bulb. When a light bulb, i.e. a tungsten light bulb, is turned on, the bulb filament generates heat. Thus, an input in the form of $\epsilon(x,t)=c+av(x,t)$ can be a model for such a problem. In here, $c$ would represent the constant source/absorption and again, the $av(x,t)$ term can represent the noisy input/output. Depending on the material property and problem investigated, the random number set, $v$, can have a nonzero mean and can attain a distribution different than the uniform distribution. These distributions can be Gaussian, Rayleigh, Rician, just to name a few.

Other possible application areas include but are not limited to turbulence, chemistry, biology, statistical finance and mathematics, nuclear reactor physics, subsurface water infiltration and nonlinear consolidation, just to name a few. That is, the diffusion of heat under the effects of nonlinearity and noisy input/output conditions can be studied in the frame of our findings. Noise would represent diverse phenomena when this broad application areas are considered. Fluctuations of the existing or input/output material concentration, heat, population of species, nuclear reactions can be modeled as the noisy inputs in the framework proposed in our paper. Thus, the rogue waves can be observed and analyzed in those broad areas of science and engineering. With these possible applications in the near future, the analysis, early detection and efficient sensing of such rogue heat (diffusion) waves will emerge as an important research area. With this motivations, spectral Fourier and wavelet analysis, Kalman filtering, quantum Kalman filtering and deep learning techniques can be effectively used to analyze, predict and early detect the rogue heat (diffusion) waves and associated supercontinuum generation.

\section{\label{sec:level1}Conclusion and Future Work}

In this paper, we have considered two different nonlinear heat (diffusion) equations and showed that noise imposed as an input triggers modulation instability, thus large amplitude fluctuations can be observed in the frame of these equations. We have showed that these fluctuations are very dynamic, thus it is possible to name them as rogue heat (diffusion) waves. We have discussed the properties and statistics of such rogue waves in the frame of the two different model equations considered. We have discussed the similarities and differences of these rogue heat (diffusion) waves from their counterparts studied in the frame of wave equations. Our findings can be used to model the effect of noise and MI on rogue wave generation in the fields including but are not limited to heat transfer, diffusion, turbulence, chemistry, biology, statistical finance and mathematics, nuclear reactor physics, subsurface water infiltration and nonlinear consolidation.

\end{document}